\documentclass{elsart}

 \usepackage{graphicx}

\def\dst{\displaystyle}

\begin{document}

\begin{frontmatter}



\title{Microcanonical and canonical approach\\ to traffic flow}


\author{Anton \v Surda}

\address{Institute of Physics SAS, 845 11 Bratislava, Slovakia}

\begin{abstract}

A system of identical cars on a single-lane road is treated as a microcanonical
and canonical ensemble. Behaviour of the cars is characterized by the 
probability of car velocity as a function of distance and velocity of the car 
ahead. The calculations a performed on a discrete 1D lattice with discrete car 
velocities.

Probability of total velocity of a group of cars as a function of density is 
calculated in microcanonical approach. For a canonical ensemble, fluctuations 
of car density as a function of total velocity is found. Phase transitions 
between free and jammed flow for large deceleration rate of cars and formation 
of queues of cars with the same velocity for low deceleration rate are 
described.

\end{abstract}

\begin{keyword}

traffic flow \sep microcanonical ensemble \sep canonical ensemble 

\PACS 05.20.Gg \sep 05.50.+g \sep 05.60.Cd \sep 89.40.Bb  
\end{keyword}
\end{frontmatter}

\section{Introduction}
\label{intr}

Traffic flow  of a system of identical cars on a single-lane road
has been intensively studied in recent decade using dynamical or 
kinetic 
description of car behaviour. [1,2]  The models used were continuous (fluid 
dynamical 
models), car-following models [3],  or discrete particle hopping models 
related to cellular 
automaton models with  stochastic behaviour [4]. In this paper we develop an 
approach to this problem based not on equations of  motion or master equations 
describing the system of cars,
but in the spirit of statistical mechanics where:  

-- Each state of the system is occupied by 
equal probability and  physical properties of the system are analyzed 
calculating number of states   for some fixed physical quantities in 
the microcanonical description. Logarithm of number of states is called 
entropy. 

-- In the canonical approach the probability of states 
depends on their energy, and the logarithm of number of all states (weighted 
by their probabilities, with negative sign and divided by temperature). 
is called free energy. 

Here we assume that
the probability of the car velocity is a function  of the velocity, and the 
distance of the car 
ahead, while all the distances between two cars are equally probable, i.e.,  
from the point of view of statistical mechanics,
a combination of the canonical and microcanonical approach is used. 
Despite of that,  further it is denoted as a microcanonical one.

The car 
distances are treated purely microcanonically only in the first part  of the 
paper. In the second one, the density of a group of cars is not fixed, but it 
is a part of larger microcanonical ensemble with  fixed number of cars. 
Here the probability of distances between two cars depends on the length of 
the group, and  the distances are treated in the same way as  velocities, 
i.e.,  canonically.

In our approach no assumptions about the drivers' behaviour and car properties 
 are necessary -- the probability of the car velocity can be measured 
experimentally, nevertheless, in this paper it is derived from a simple model 
behaviour of cars. 
 
In statistical physics the term microcanonical ensemble means, as a rule, an 
isolated system, in which some physical quantities are conserved and thus are 
constant. In our microcanonical approach the system 
is not assumed to be isolated, but  
only such states of the system, or a subsystem, are taken into account, for 
which some quantities remain 
constant. In a  system of cars this may be, e.g., sum of the velocities of 
all cars or their density.   The subsystem is influenced by a boundary 
condition -- the distribution of velocities of the car ahead of the 
investigated group.

In the approach of Mahnke et al. [5]  the group of cars is represented by a 
grandcanonical 
ensemble,  number of cars in which is not fixed, and its chemical potential 
is a function of parameters of a master equation. In our approach the 
thermodynamics of the system of cars is systematically derived starting 
from the microcanonical approach. The properties of a canonical 
ensemble are deduced from the known entropy of a subsystem together with a 
reservoir.
When the size of the cluster of cars inside the reservoir is changed, the 
derivative of entropy represents a pressure  exerted by reservoir of cars 
on the group, instead of chemical potential in the above-mentioned work.

In the last years we could observe  a revival of the microcanonical approach to
the problems of statistical mechanics [6--10].  One of the reasons for it was 
that the 
region where the entropy of a finite system is convex, instead of the standard 
concave 
shape of it,  was identified as a point or line in the phase space where the 
first-order phase transition in corresponding infinite system takes place.  
As the number of observed cars in normal traffic is not too large, the 
techniques developed in statistical physics for small systems are convenient 
in this case. The term ``phase transition'' in this paper is used in the sense 
of the above-cited works.

\section{Model and method} 
\label{model} 

The cars are further represented by dimensionless points moving on a discrete 
one-dimensional lattice, and are characterized by 2 quantities:  discrete 
velocity $v_i$ in the interval $\langle0, v_{\rm max}\rangle$ and a discrete 
coordinate (site number)   $x_{i} \in 
\langle1, L\rangle$. 
$v_{\rm max}$ is the maximum  velocity given by the construction of the car  
and $L$ is the length 
of the observed group (subsystem)  of cars. The coordinate of each car is 
measured with respect to the last car of the group. Its coordinate is 
always 0, i.e.,  the origin of the coordinate system is fixed to it. As the 
length of the group is $L$, the coordinate of the 
last car of the group ahead is $L$. Number of cars in  the group is $N$. (The 
lattice constant is related to the car length).
Car velocities and coordinates acquire only integer values.
 Car velocities are random, described by a probability distribution
peaked around an optimal velocity  $v_{\rm opt}$,
which  is further chosen as 90\% of maximal safe velocity $v_{\rm m}$.
The maximal safe velocity is determined from the requirement that two 
neighbouring cars, which start to decelerate at the same time with the  same 
deceleration rate $a$, would stop without crash. Moreover,   $v_{\rm m}$ must 
not be greater than the maximum possible velocity of the car  $v_{\rm max}$, 
i.e.,
 for every car 
\begin{eqnarray} 
\label{eq1} 
& & v_{\rm opt}(v_2,x_{1,2}) = 0.9 v_{\rm m}(v_2,x_{1,2}) ,\nonumber\\ 
& &  v_{\rm m}(v_2,x_{1,2})=
\left\{ {\begin{array}{*{20}l}
\sqrt{2ax_{1,2}+v_2^2}& {\rm if}\  v_{\rm m}\le v_{\rm max}\\
v_{\rm max}\quad & {\rm if}\  v_{\rm m}> v_{\rm max}\\ 
\end{array}}\right.  
\end{eqnarray} 
 where $x_{1,2}$ and $v_2$ are the distance (headway) and velocity of the car 
ahead, respectively. The reaction time of the driver in (1)  is assumed to be 
equal to zero, 
nevertheless, it can be easily generalized for nonzero reaction times with 
only a small impact on our final results. (This problem is discussed in more 
detail in [11].   As we use only integer values of 
velocities, the nearest integer value to $v_{\rm opt}$ from (1)  is taken    
for the actual optimal velocity in our calculations.

The way of driving of the observed drivers is characterized by 
 distribution of probabilities of car velocities around the optimal velocity.
Here we use an extremely simple distribution, in which the probability of 
optimal velocity is $p_0$, the probabilities of the velocities $ v_{\rm opt}\pm
1$ are $p_1$, while the probability of the car to have any other 
permitted velocity is $p_2$. The sum of all probabilities for each car is 
equal to 1. The parameters $p_0, p_1$ and $p_2$ are the same for every
car, 
and the distribution depends on the headway only by means of the value of 
optimal velocity.

\section{Microcanonical description}

In the microcanonical approach  only such groups of $N$ cars, which 
length is $L$ and sum of their velocities is $V$, are studied. These groups of 
cars are 
 influenced only by the velocity distribution of the car ahead of them with 
coordinate $L$. The probability distribution of each car is given by the rule
above as a function of headway and the velocity of the car ahead, while the 
distances between them are arbitrary and limited only by the length of 
the group. 

The probability that the sum of  velocities of $N$ cars in a 
group 
of length $L$ is $V$ multiplied by the number of their configurations on $L$ 
sites, is further denoted as
$W(V,L)$ and called density of states.
It can be calculated recurrently 
\begin{eqnarray} 
\label{eq2} 
& & W_1(V_1, L_1; v_2) =p(V_1; L_1,v_2)\nonumber\\ 
& & \vdots\nonumber\\
& & W_i(V_i, L_i; v_{i+1}) =\sum_{v_i, x_{i,i+1}}
W_{i-1}(V_i-v_i, L_i-x_{i,i+1}; v_i) p(v_i; x_{i,i+1},v_{i+1})\nonumber\\
& & \hbox{for } i=2,N-1\nonumber\\ 
& & \vdots\\
& & W_N(V, L; v_{N+1}) =\nonumber\\
& &\qquad {} =\sum_{v_N,  x_{N,N+1}}
W_{N-1}(V-v_N, L-x_{N,N+1}; v_N) p(v_N; x_{N,N+1},v_{N+1})\nonumber\\
& & W(V, L) =\sum_{v_{N+1}} W_{N}(V, L; v_{N+1}) p(v_{N+1})\nonumber
\end{eqnarray} 
where $0\le v_j \le v_{\rm max}$, $0\le V_j \le j v_{\rm max}$,  
$j\le x_{j,j+1}, L_j\le L - j$. $p(v_{N+1})$ in the last line of (2)  is the 
velocity probability of the last car of a large group ahead of the studied 
group   with the same car density. This large group will be further called 
reservoir.

Density of states  in the reservoir of length $L_{\rm r}$, number of 
cars $N_{\rm r}$, with the density $\dst{ N_{\rm r}\over L_{\rm r}}= {N\over 
L}$, and fixed velocity of the last car $v_{N+1}$ is   
\begin{equation}
\label{eq3} 
W_{\rm r}(v_{N+1},L_{\rm r})   =  \sum_{v_{N+2},\dots,v_{N+N_{\rm r}}\atop
x_{N+1,N+2},\dots,x_{N+N_{\rm r},N+N_{\rm r}+1}}     
\prod_{i=N+1}^{N+N_{\rm r}}   p_i(v_i; x_{i,i+1}, v_{i+1})\delta_{L_{\rm 
r},\sum x_{i,i+1}}
\end{equation}
It depends, in principle, on the velocity of the first car of the reservoir, 
but  numerical calculations show that for large $N_{\rm r}$ this dependence 
is
negligible. The probability $p(v_{N+1})$ is the normalized  density of states
$W$
\begin{equation}
\label{eq4} 
p(v_{N+1})= W_{\rm r}(v_{N+1},L_{\rm r})/\sum_{v_{N+1}}W_{\rm 
r}(v_{N+1},L_{\rm r}) 
\end{equation}

The quantity $W(V,L) $ in (\ref{eq2})  expresses the probability that the sum 
of velocities of the 
cars in the group is $V$ as well as the number of possible configurations of 
occupation of $L$ sites by $N$ cars. As  mentioned above, it is, in fact,  a 
product of probability $P$ and number of configurations $\Omega $
\begin{equation}
\label{eq5} 
W(V,L) ={W(V,L)\over\sum_V W(V,L)}\cdot \sum_V W(V,L) \equiv P(V,L)\cdot  
\Omega(L).  
\end{equation}

As for the fixed length of the subsystem, $\Omega(L)$ is constant, only the 
normalized probability  $P(V,L)$ is be presented in  Results.

In the microcanonical approach  only subsystems of cars with the constant 
density, the same  as is the mean density of the whole  system, are studied. 
To take into account 
also the density fluctuations, it is more convenient to use the canonical 
description with variable density of the subsystem due to its variable length.

\section{Canonical description}

In canonical approach the length of the subsystem varies,
 only the length of the whole system, subsystem + 
reservoir  is fixed. The number of cars in the subsystem and in the reservoir 
remains constant, so the density of cars varies with varying length of the 
groups. 
Our canonical description differs from the grandcanonical approach of Mahnke et
al.
[5]  where  the density of the 
subsystem changes due to exchange of cars between the subsystem and reservoir.

In statistical mechanics the properties of a  reservoir are usually not 
calculated, 
only the values of derivatives of its entropy (logarithm of number of states)  
with respect to the quantities, which are fixed in the whole system, are  
assumed to be known. They are, e.g., temperature, chemical potential, etc. 
Similarly, in our
 canonical description of the  system of cars, a pressure of  reservoir 
exerted on the 
subsystem could be introduced. Nevertheless, this quantity cannot be directly 
measured, and it would depend on the velocity of the last car of the 
reservoir, so we prefer a direct calculation of number of states of a large 
enough reservoir for given velocity of the last car and  length of the 
reservoir. 
 
The length of the system $L_{\rm s}$ is the sum of the length of the subsystem 
and 
reservoir $X+L_{\rm r}$. The number of cars in the  subsystem and 
reservoir  are 
fixed and  denoted as $N$ and $N_{\rm r}$, respectively. If $X\ll L_{\rm r}$
and $N\ll N_{\rm r}$, the properties of the subsystem does not depend on 
velocity of the first car in the reservoir. 
 
Density of states of the reservoir at given velocity of the last car  is 
calculated according to ({\ref{eq3}). 
Density of states of the whole system at given total velocity of the subsystem 
$V$ and its  length $X$ can be obtained by the same way  as in the 
microcanonical case, only in the last term in ({\ref{eq2}) -- the probability 
of the velocity of the last reservoir car $p(v_{n+1})$ -- is replaced by the  
density of 
states of the reservoir. Last line of ({\ref{eq2}) now reads
\begin{equation}
\label{eq6} 
W(V, X) =\sum_{v_{n+1}} W_{n}(V, X; v_{n+1}) W_{\rm r} (v_{n+1},L_{\rm s} - 
X). 
\end{equation}

The mean density of the subsystem $N/\langle X\rangle$ is equal to the density 
of the whole system $(N+N_{\rm r})/  L_{\rm s}$.

The main difference between the microcanonical and canonical treatment is that 
in the first case only number of states of the subsystem is calculated  while 
in the latter case the properties of the subsystem  are given by the number of 
states of the whole system. In the microcanonical approach the reservoir is 
used 
only for calculation of boundary condition -- probability distribution
of the last car of the  subsystem.
It is summed 
 over all car velocities and positions. 
In the canonical system the summation is 
performed 
only over velocity degrees of freedom as the velocity even of the whole 
system is not conserved. 
 
\section{Results and discussion} 

The velocities and positions of cars are described by discrete variables in our
model. Changing the values of its parameters, we can observe two 
different types of behaviour. In the first one,  for high deceleration 
rate $a$ and low densities, the system behaves like continuous;  in the 
density 
of states the underlying discrete structure of velocities is not seen. At 
small $a$ and high densities,  total velocities of the system, which  are 
integer multiples of number of particles, are more probable then the others.  
This regime reminds a ferromagnetic Potts model where the total 
magnetization of the system points in many different directions of the space.

In all our calculations, car velocity acquires 21 values $v_i=i,\ i=0,20$. The 
probability of a car to move with a velocity $v_i$ depends on the velocity and 
distance of the car ahead by means of optimal velocity $v_{\rm opt}(v,x)$.  
It acquires 3 values $p(v_{\rm opt})= p_0$, $p(v_{\rm  opt}\pm1)= p_1$, 
$p(v_{i})=p_2$ for all other $v_i$. Only two of these parameters are 
independent as the probability is normalized:  $\sum_{i=0} ^{20}  p(v_i) 
=1$. In the present calculations $p_1$ is fixed to 0.3, and $p_2$ was chosen  
for 
the only free parameter of the velocity distribution.   The position of a car 
with respect to the 
first one may be an integer between 0, and $L-1$ if the site is not occupied 
by another car. In the free-flow regime the deceleration rate $a$ in (1) is 
put equal to 4.0, in the jammed regime, where the discreteness of the velocity 
plays role, $a=0.5$.

The main result of our calculations are density of states $W(V,X)$ in 
the canonical case and probability of the total velocity, $P(V,X)$, of the 
subsystem as a 
function of the subsystem length at fixed number of cars for microcanonical 
ensemble. 
 They are plotted in 3D graphs.

\begin{figure}
\begin{center}
\includegraphics*[width=13.5cm]{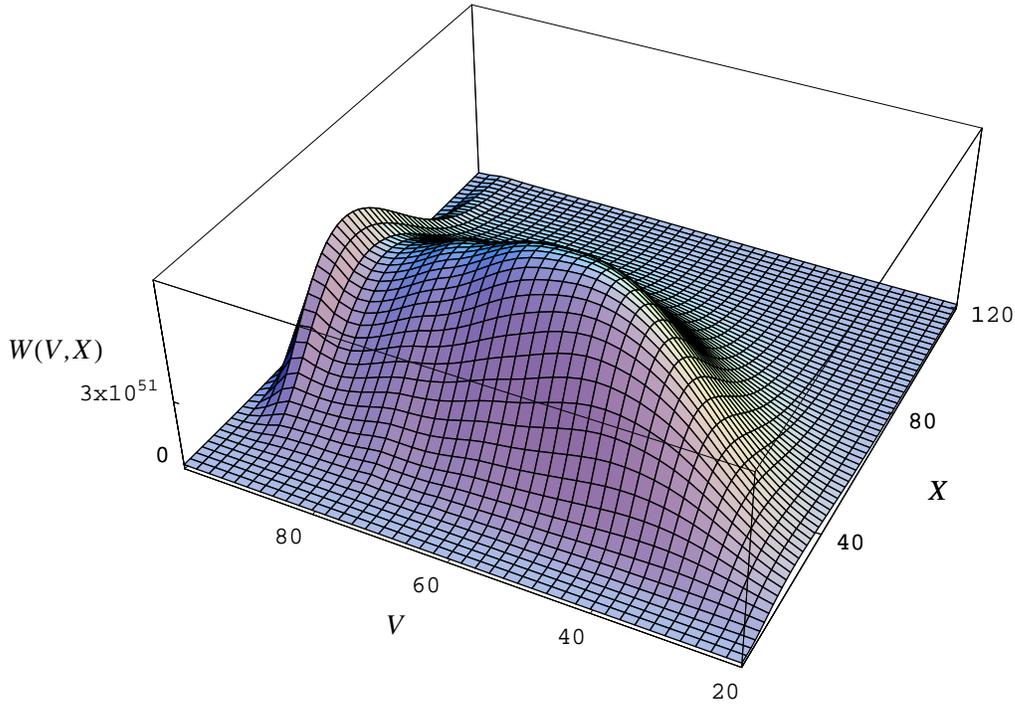}
\end{center}
\caption{Density of states of a canonical ensemble of 5 cars as a 
function of its 
total velocity $V$ and length $X$ for $a=4.0$, $p_2= 0.03$, and mean length 
$\langle X\rangle=50$, i.e., mean density $\rho=0.1$.} 
\label{fig:1}
\end{figure}

\begin{figure}
\begin{center}
\includegraphics*[width=13.5cm]{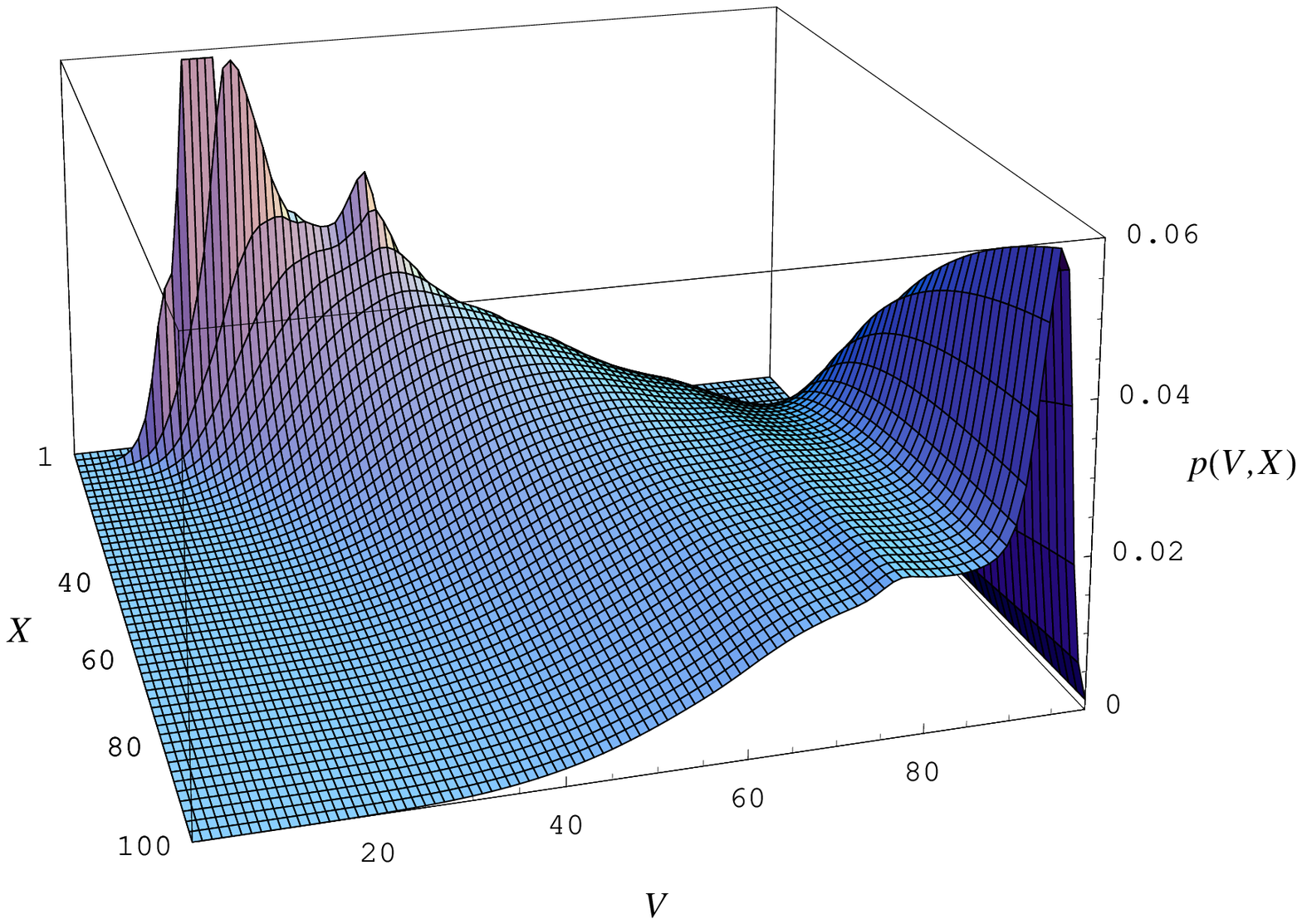}
\end{center}
\caption{Probability of the total velocity  of a microcanonical ensemble 
of 5 cars as a function of its  total velocity $V$ and length $X$ for $a=4.0$, 
$p_2= 0.025$. The density of the system varies from 0.05 to 1.} 
\label{fig:2}
\end{figure}

\begin{figure}
\begin{center}
\includegraphics*[width=13.5cm]{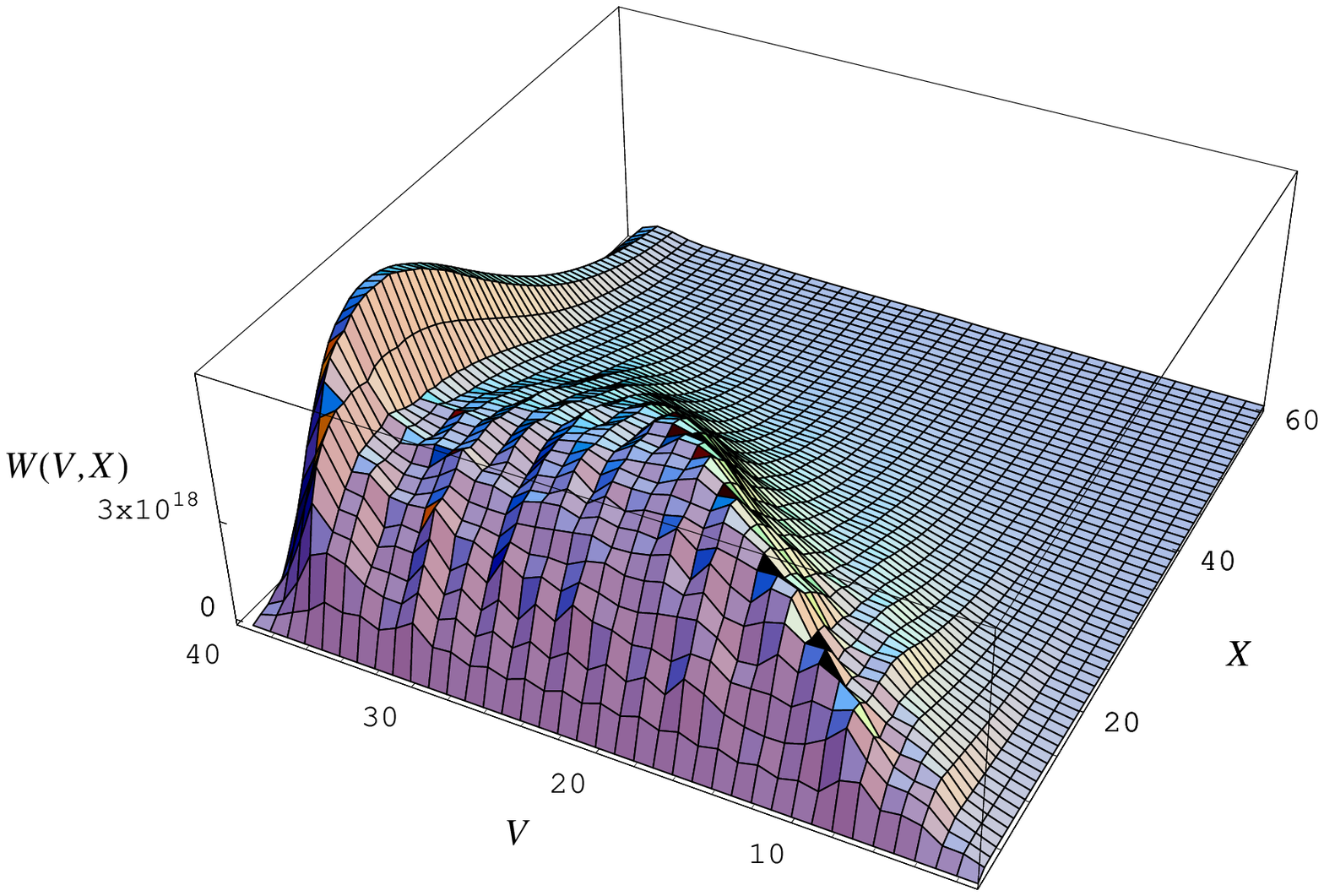}
\end{center}
\caption{Density of states of a canonical ensemble of 2 cars as a function of 
its total velocity $V$ and length $X$ for $a=4.0$, $p_2= 0.03$ and mean length 
$\langle X\rangle=25$, i.e., mean density $\rho=0.1$.\vspace{-0.6cm}} 
\label{fig:3}
\end{figure}

\begin{figure}
\begin{center}
\includegraphics*[width=11.5cm]{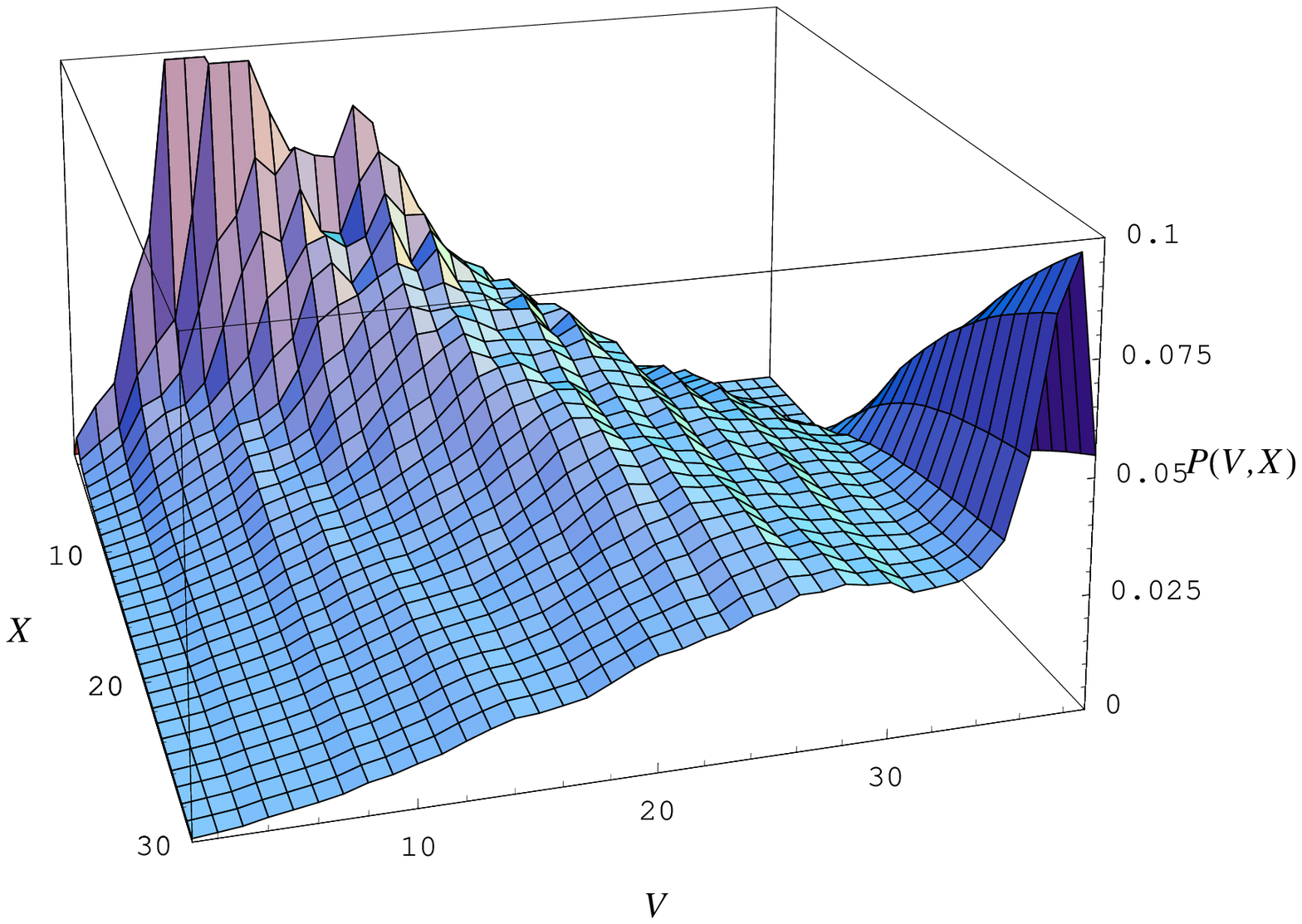}
\end{center}
\caption{Probability of the total velocity  of a microcanonical ensemble of 2 
cars as a function 
of its total velocity $V$ and length $X$ for $a=4.0$, $p_2= 0.03$. The
 density of the system varies from 0.05 to 1.\vspace{-0.6cm}} 
\label{fig:4}
\end{figure}

In the free-flow regime with $a=4$,  the shape of density of states $W$, in 
canonical ensemble, depends on  the parameter $p_2$. For $p=0$, practically 
all the cars have their velocity close to the optimal one with the most 
probable 
density at the value of the mean density of the system. For large 
probabilities 
of the small and large velocities, $p_2>0.05$, the density of states is 
represented 
by a broad peak with maximum of less than one half of the optimal velocity, 
and the cars also in this regime become jammed.  In Fig.~1 an intermediate 
case is shown 
with a narrow free-flow peak and a broad peak of jammed cars. The plot 
represents distribution of density of states  for 5 cars creating a group of 
length $X$ with 
total velocity $V$. The group is a part of a large system of cars with fixed 
total length with density 0.1, i.e., the mean length of the group of 5 cars 
is 50.  

As expected, the velocity of jammed cars is lower that of those moving freely. 
Most probable total velocity of the jammed group is about 2/3 of the total 
velocity 
  of freely moved group, which density is about 20\% smaller. As the car flow 
in both 
groups is different, they cannot coexist in a finite system in steady state.   
Both peaks represent two states of the system of cars with very low 
transition rate between them, which is expressed by the depth of the minimum 
between them.
Two neighbouring fluctuation can live long only if 
they have the same mean velocity per particle. On the other hand, as the plot 
has only one maximum for constant  $V$, the probability of such 
fluctuations is low.

In the microcanonical case the length of the subsystem determines the 
density of cars not only in the subsystem but also in the whole system 
together with reservoir.  The length of the group is fixed, and there are no 
length 
fluctuation of it.   The probability of total velocity of 5 cars for various  
density of the whole system is shown in Fig.~2. The plot is viewed from 
opposite direction than in Fig. 1. For length of the group 
$X=50$, it corresponds to the canonical system of length $L= 50$ (Fig.~1), 
which 
is equal to the mean length of the group. While in Fig.~1 the plot for small 
$X$ represents behaviour of dense fluctuations, in Fig.~2 the whole system 
together 
with reservoir is dense. It can be seen that for $X=5$ the most probable state 
is when all cars have velocity 4.

For high densities the most probable state has a small velocity, for low 
densities the most probable velocity is the optimal one. The first-order phase 
transition occurs when the height of low and high velocity peak is equal; in 
Fig.~2, it is at $X=38.$ It should be stressed that the phase transition 
introduced in this paper is not the  phase transition exactly in 
thermodynamic 
sense as the system of cars is finite.  In this system  cannot coexist 
large 
groups of cars in different phases, i.e., with different velocities. Then the 
transition from a local maximum to absolute maximum of probability is very 
improbable, and a strong hysteresis occurs in the system, observed also 
experimentally [12].

In Fig.~3 and 4 the same plots, but  for 2 cars only, are shown. The free-flow 
peak is  more pronounced in Fig. 3. It can be explained by the fact that also 
in a jammed 
5-car group,  2 cars may move fast for a short time. The average length of 
the group in canonical ensemble is now 20.  For two-car groups the discreteness
of model velocities in the density of states and probability of velocity is 
manifested. For 5 or more cars, these quantities are smoothed, and the model 
can  simulate to some extent a real traffic.

The discreteness is conspicuous for low braking ability of cars and high 
densities.
For $a=0.5$ and  $N/\langle X\rangle=0.2$ it can be seen
 in Fig.~5. In this case, 
especially for fluctuation with high densities, the cars have tendency to form 
queues with the same velocity of each car.

\begin{figure}
\begin{center}
\includegraphics*[width=11.5cm]{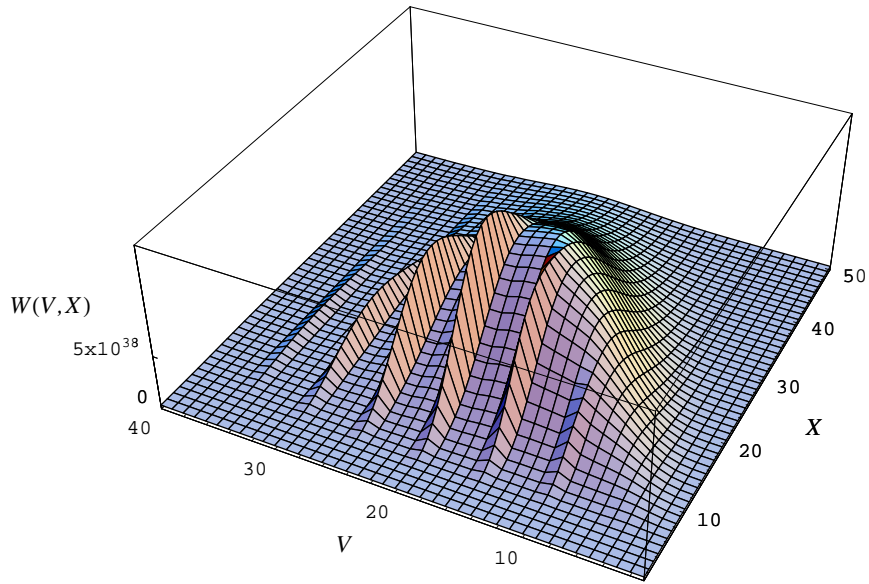}
\end{center}
\caption{Density of states of a canonical ensemble of 5 cars as a 
function 
of its total velocity $V$ and length $X$ for $a=0.5$, $p_2= 0.01$ and mean 
length $\langle X\rangle=25$, i.e., mean density $\rho=0.1$.\vspace{-0.5cm}}   
\label{fig:5}
\end{figure}

\begin{figure}
\begin{center}
\includegraphics*[width=11.5cm]{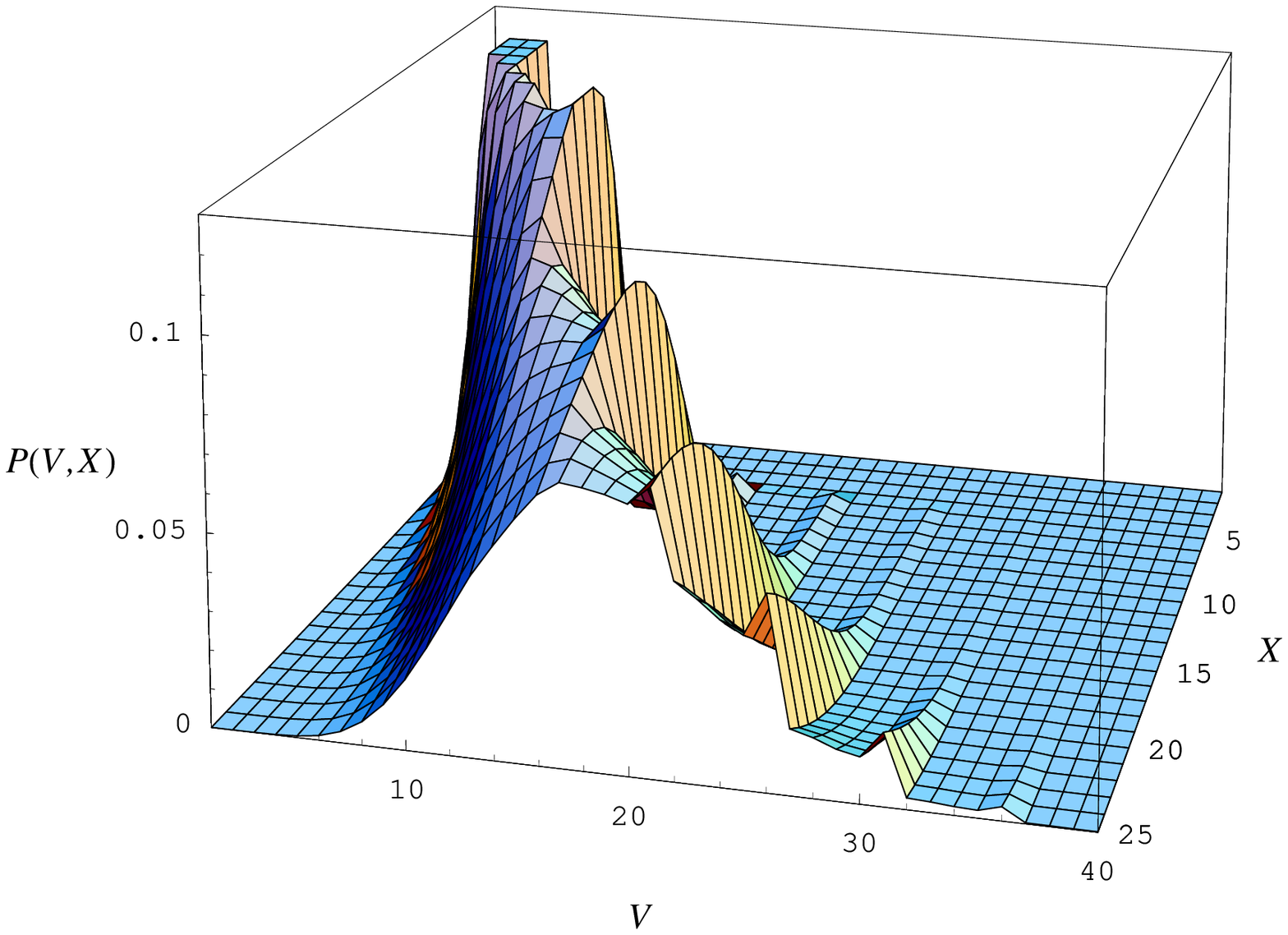}
\end{center}
\caption
{Probability the of total velocity  of a microcanonical ensemble 
of 5 cars as a function of its total velocity $V$ and length $X$ for $a=0.5$, 
$p_2= 0.01$. The density of the system varies from 0.2 to 1.} 
\label{fig:6}
\end{figure}

\begin{figure}
\begin{center}
\includegraphics*[width=11.5cm]{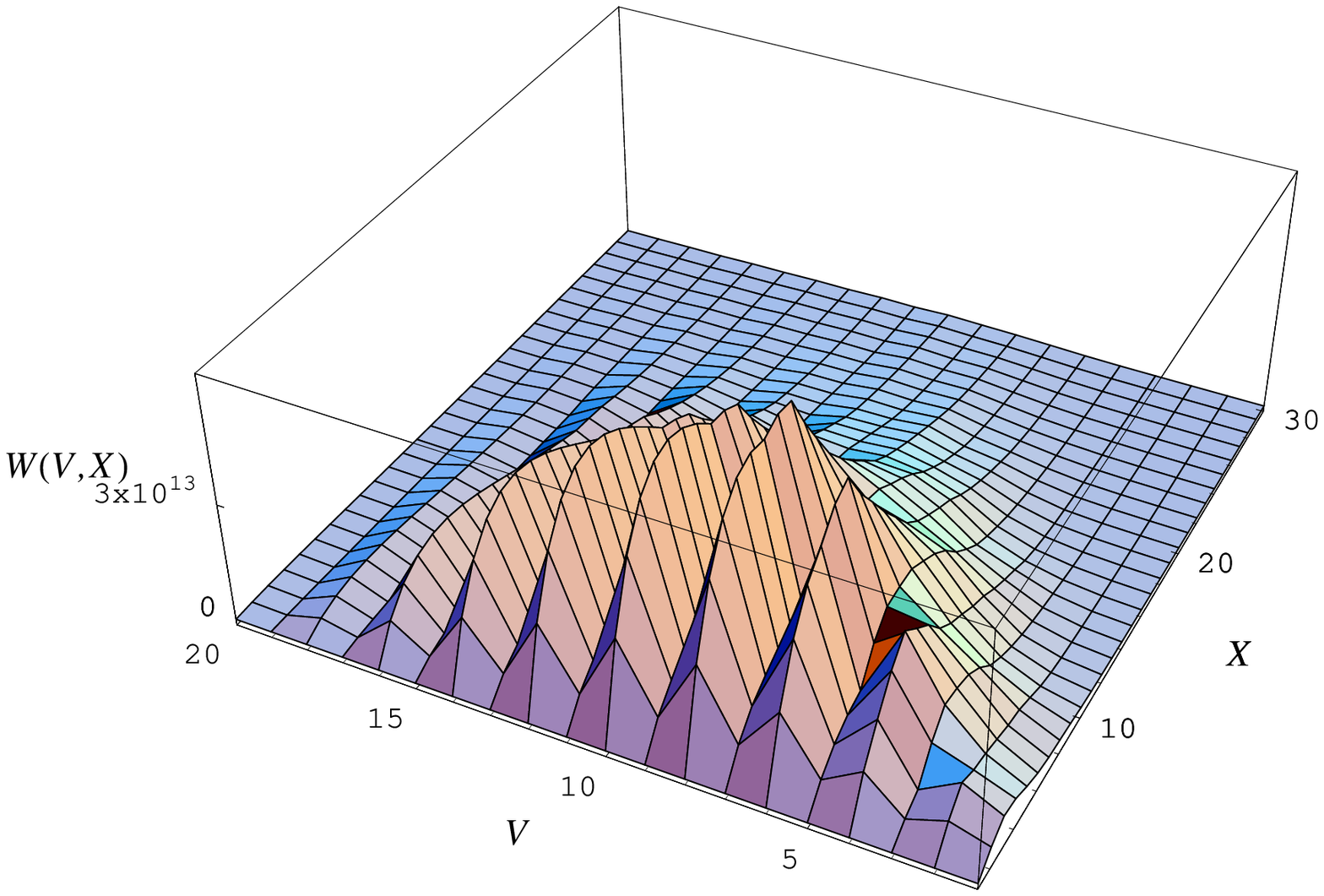}
\end{center}
\caption{Density of states of a canonical ensemble of 2 cars as a 
function 
of its total velocity $V$ and length $X$ for $a=0.5$, $p_2= 0.01$ and mean 
length $\langle X\rangle=10$, i.e., mean density $\rho=0.2$.\vspace{-1cm}} 
\label{fig:7}
\end{figure}

\begin{figure}
\begin{center}
\includegraphics*[width=11.5cm]{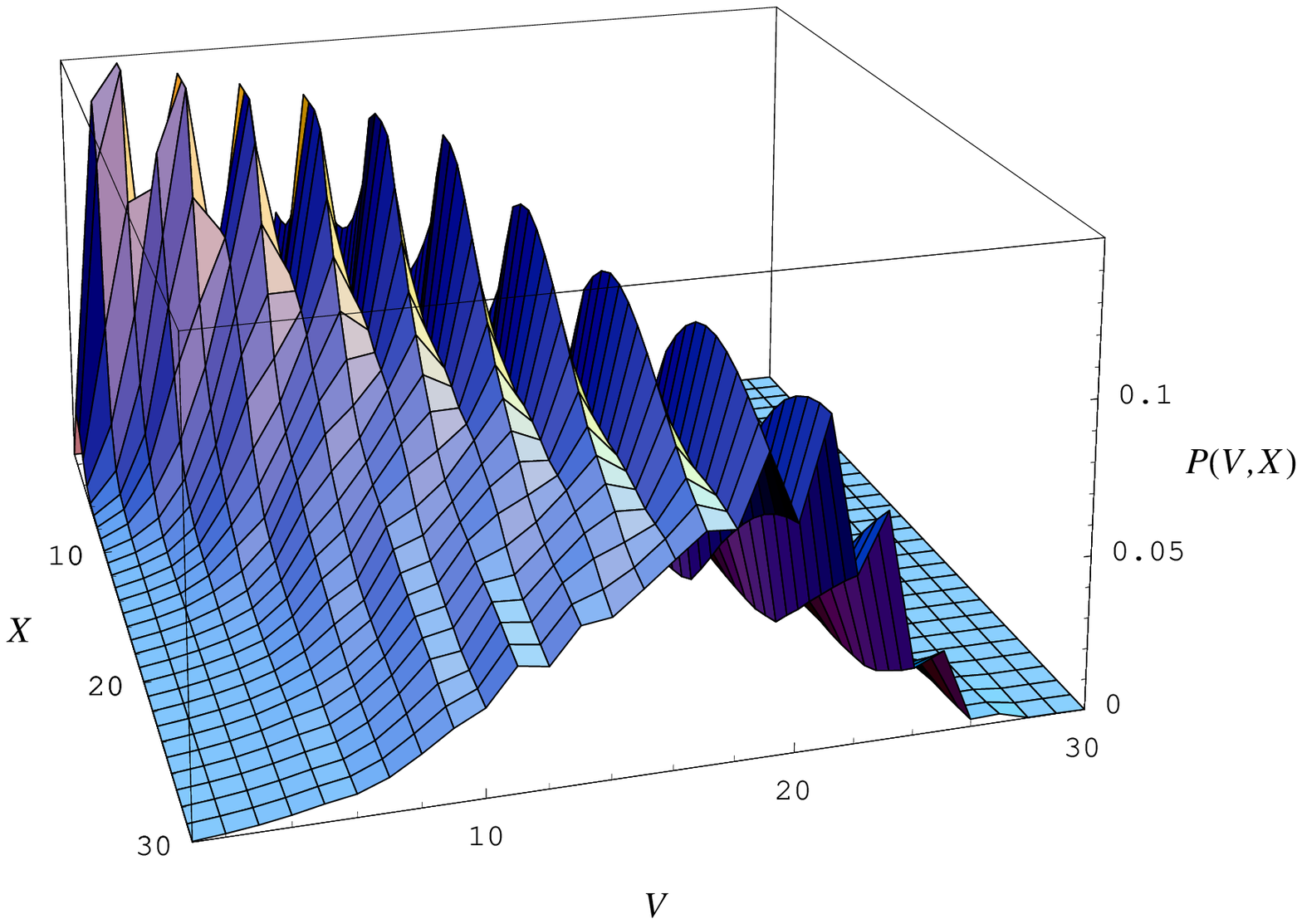}
\end{center}
\caption
{Probability of the total velocity  of a microcanonical ensemble 
of 2 cars as a function of its total velocity $V$ and length $X$ for $a=0.5$, 
$p_2= 0.01$. The density of the system varies from 0.066 to 1.\vspace{-0.6cm}} 
 \label{fig:8}
\end{figure}

For low densities $(X>25) $ and 
small velocities, these queues are dissolved. The microcanonical picture of 
the system is in Fig.~6. Now probability distributions have a number of local 
maxima. Each of them can become an absolute maximum for some values of $X$. 
Large velocity maxima are absolute maxima for low densities of cars.

For a system of 2 cars with the same parameters, the 
density of states and probabilities are in Figs. 7, 8. Here the queues of cars 
with equal velocity consist 
of 2 cars, and the probability of  total velocity has peaks at even values 
of $V$.

For small deceleration rates $a$,  the  velocity probability consists of 
number of peaks representing phases with different total velocities and 
first-order phase transitions between them. The difference between these 
velocities 
can be taken as an order parameter in the concrete phase transitions. With 
increasing $a$ the valley between two probability peaks is disappearing, the 
order 
parameter becomes zero, and a second-order phase transitions takes place. For 
large $a$ (Fig. 1, 2) the distance between the peaks remains always large.  
Changing the 
parameter $p_2$, one of the peaks disappears, but merging of two peaks into 
one, i.e. the second-order phase transition,  is never observed. 
 
In conclusion, the microcanonical and canonical decription of a system of cars 
was developed. The only input into the theory is  
 the 
probability of car velocity as a function distance and velocity of the car 
ahead. 
According to standard procedure in  statistical mechanics, all other missing 
information are replaced by the principle of maximum entropy 
of the system. From these assumptions pressure-density diagram of the system 
can 
be derived, but here, only directly observable quantities as density of states 
and  probability of the total velocity  of a group of cars were presented.

We acknowledge support from VEGA grant No. 2/6071/2006.

\end{document}